\documentstyle[12pt,aasms4]{article}

\slugcomment{Submitted to Astrophysical Journal (Letters).}

\lefthead{Scoville et al.}
\righthead{CO (J = 3 - 2) Emission in the Radio Galaxy 53W002 at z=2.394}

\newbox\grsign \setbox\grsign=\hbox{$>$} \newdimen\grdimen 
\grdimen=\ht\grsign
\newbox\laxbox \newbox\gaxbox
\setbox\gaxbox=\hbox{\raise.5ex\hbox{$>$}\llap
     {\lower.5ex\hbox{$\sim$}}}\ht1=\grdimen\dp1=0pt
\setbox\laxbox=\hbox{\raise.5ex\hbox{$<$}\llap
     {\lower.5ex\hbox{$\sim$}}}\ht2=\grdimen\dp2=0pt

\def\kms    {\ifmmode{{\rm ~km~s}^{-1}}\else{~km~s$^{-1}$}\fi}

\def\Mo     {~$M_{\odot}$}

\def\Msun     {~$M_{\odot}$}
\def\deg      {{\ifmmode^\circ\else$^\circ$\fi} } 
\def\Ho       {{$H_{0}$} }
\def\qo       {{$q_{0}$} }
\def\kmsMpc   {{\ km\ s$^{-1}$\ Mpc$^{-1}$} }
\def\h2     {H$_2$}
\def\etal   {{\sl et~al.~}}
\newbox\grsign \setbox\grsign=\hbox{$>$} \newdimen\grdimen 
\grdimen=\ht\grsign
\newbox\simlessbox \newbox\simgreatbox
\setbox\simgreatbox=\hbox{\raise.5ex\hbox{$>$}\llap
     {\lower.5ex\hbox{$\sim$}}}\ht1=\grdimen\dp1=0pt
\setbox\simlessbox=\hbox{\raise.5ex\hbox{$<$}\llap
     {\lower.5ex\hbox{$\sim$}}}\ht2=\grdimen\dp2=0pt

\begin{document}

\title{CO (J = 3 -- 2) Emission in the Radio
Galaxy 53W002 at z=2.394}

\author{N. Z. Scoville}
\affil{California Institute of Technology, Pasadena, CA 91125}

\author{M. S. Yun}
\affil{National Radio Astronomy Observatory,
	P.O. Box 0, Socorro, NM 87801}

\author{R. A. Windhorst}
\affil{Department of Physics and Astronomy, Arizona State University, 
Box 871504, \\ Tempe, AZ 85287-1504}

\author{W. C. Keel}
\affil{Department of Physics and Astronomy, University of Alabama, 
Box 870324, \\ Tuscaloosa, AL 35487-0324}
\and 

\author{L. Armus}
\affil{California Institute of Technology, Pasadena, CA 91125}


 

\begin{abstract}
We report a sensitive search for redshifted CO (3--2) emission from the
weak radio galaxy 53W002 at z=2.394.  Maps at resolutions of 3\arcsec\ and
235\kms\ show a significant emission peak within 0.5\arcsec\ of the
optical and radio continuum peaks. The measured narrow band flux is
approximately ten times the extrapolated cm-wavelength non-thermal radio
continuum expected at 101.9 GHz and exhibits a spectral profile implying a
540\kms\ width (FWHM) at a systemic redshift z $= 2.394\pm0.001$ for CO
(3-2). This emission has a total integrated flux of $1.51\pm0.2$ Jy \kms,
approximately 4 times weaker than that previously seen in the lensed
systems FSC10214+4724 and the Cloverleaf QSO.  For a Galactic CO-to-H$_2$
conversion ratio, the implied molecular gas mass is
7.4$\times$10$^{10}$\Msun ~(\Ho=75\kmsMpc and \qo=0.5).  The CO emission
is elongated at PA = 120\deg with a deconvolved major axis radius of 15
kpc (2.8\arcsec). This extension is along a similar direction to that seen
in the cm-wave radio continuum and the optical but approximately three
times larger.  A velocity gradient is seen along the major axis, and if
this structure is a (forming) disk, the implied dynamical mass is
9-22$\times$10$^{10}$ \Msun\ at r $\leq 15$ kpc, assuming inclination i =
0\deg (edge-on). The magnitude of these masses and the similarity of the
high gas-mass fraction are consistent with the host galaxy of 53W002 being
a young galactic system but the metallicity (probably $\ge 0.1 Z_{\odot}$
in order to produce the CO lines) implies significant heavy element
production prior to z=2.4.  This constitutes the first high redshift
molecular gas which is detected in emission where there is probably no
gravitational magnification.
\end{abstract}


\section{Introduction}

Early in the evolution of galaxies, there will be a phase in which the
predominant mass component is interstellar and observations of this matter
are critical to progress in understanding the initial building of galaxies
and their stellar populations. Although the protogalactic gas should
initially be atomic (or ionic), virtually all star formation models
suggest that the gas becomes molecular during collapse to form stars. With
relatively small metal enrichment ($\geq 10\%$ solar metallicity) in a
prior, low-level starburst, this denser ISM becomes observable in the CO
rotational transitions at mm-wavelengths.  To date, there have been three
confirmed detections of CO emission at redshift $z \ge 1$:  FSC10214+4724
at z=2.28 (\cite{bv92,sol92a,sco95}); the Cloverleaf QSO at z=2.56
(H1413+117 -- \cite{bar94,yun97}); and BR1202-0725 at z=4.69
(\cite{oh96,om96}). In the first two (and possibly the last also, c.f.
\cite{om96}), the high redshift system is gravitationally lensed by a
foreground galaxy or cluster and the intrinsic (non-amplified) properties
are therefore too uncertain for broad inferences about protogalactic
evolution.

In this letter we report high resolution aperture synthesis observations
designed to detect and image the molecular emission in the radio galaxy
53W002 and its associated galaxy group or cluster at $z_{opt} = 2.390$
(Windhorst \etal 1991, hereafter W91; Windhorst \etal 1994; Pascarelle
\etal 1996a, 1996b). This cluster is clearly not subject to significant
gravitational lensing based on the optical morphology and the fact that
the separate galaxies have somewhat different redshifts and quite
different spectra. 53W002 is unlikely to be lensed by the lower-luminosity
z=0.581 elliptical galaxy that appears 5\arcsec\ to its North-West (c.f.,
W91; Windhorst \etal 1992, 1994).  Moreover, no other lensed images at
z=2.39 are seen immediately surrounding this object in the narrow-band
redshifted Ly$\alpha$ WFPC2 images of Pascarelle \etal (1996b).  This
radio-selected cluster was chosen for a sensitive search for molecular
emission since it appears to be a group of young, presumably gas-rich
objects and Yamada \etal (1995) have reported a possible CO (1-0) emission
feature here. Our observations clearly detect CO (3-2) emission (albeit at
a level corresponding to a mass which is probably 5--14 times less than
would be consistent with Yamada \etal).  
Throughout this letter, we adopt \Ho=75\kmsMpc and
\qo=0.5.

\section{Observations}

53W002 was observed using the Owens Valley Millimeter Array in three
configurations between February 1996 and January 1997. The array consists
of six 10.4 m telescopes, and the longest baseline observed was 242 m.
For 53W002 whose systemic optical redshift is z=2.390 (see W91), the CO
(3--2) transition occurs at 101.95 GHz. The synthesized beams were
2.39$\times 1.94^{\prime\prime}$ (uniform weighted) and 4.09$\times
2.96^{\prime\prime}$ (natural), both at PA={$-$68$\deg$}. The phase
center, central redshift for the measured CO line, and adopted angular
size and luminosity distances for 53W002 are given in Table 1. The primary
beam of the 10.4 m telescopes is 70\arcsec\ and thus the observations
cover all objects in the cluster within $\sim$35\arcsec, corresponding to
150 kpc at 53W002. Several other z=2.4 candidates are included in this
area : 5, 11, etc. of Pascarelle \etal (1996a, b).  The system
temperatures were typically 250--350 K in the signal side-band corrected
for antenna and atmospheric losses and for atmospheric incoherence.
Spectral resolution was provided by a digital correlator configured with
120$\times$4 MHz channels (11.4 \kms), covering a total velocity range of
1,300 km s$^{-1}$.  In addition, the continuum emission was measured in a
1 GHz bandwidth analog correlator.  The total useable integration time on
source was 61 hours distributed over 15 tracks in the three different
telescope configurations.  The nearby quasar 1828+487 (1.95 Jy at 102 GHz)
was used to track the phase and gain variations, and Uranus (T$_b$=125 K)
and 3C 454.3 were observed for absolute flux calibration. The positional
accuracy of the resulting maps is $\sim 0.3^{\prime\prime}$.  The data
were calibrated using the standard Owens Valley array program mma
(\cite{sco92}) and mapped using DIFMAP (\cite{she94}) and the NRAO AIPS
package. For the continuum maps (BW = 1 GHz), the rms noise was 0.20 and
0.48 mJy for natural and uniform weighting, respectively.

In order to predict more accurately the expected non-thermal radio
continuum at 101.9 GHz, we also obtained 25 min of data with the NRAO VLA
at 14.9 GHz in the B-array on March 11, 1997.  3C 286 and 1658+476 were
used for flux and gain calibration.  The synthesized beam is $0.81 \times
0.53$ \arcsec\ (PA=--82\deg) for these data.  The source was unresolved
($\leq0.31$\arcsec\ in diameter) with an integrated flux of $2.8\pm0.1$
mJy.

\section{Results}

In Figure 1 the interferometric spectrum integrated over the central
5\arcsec\ centered on 53W002 is shown, smoothed to a velocity resolution
of 235 km s$^{-1}$ and sampled every half resolution element.  The peak
flux is $3.1\pm0.5$ mJy and the deconvolved line-width is $540\pm100$ km
s$^{-1}$ (FWHM). For comparison, the peak flux of FSC10214+4724 is
$14\pm2$ mJy and the full linewidth is 250 \kms (\cite{sco95},
\cite{dow95}) and similar parameters hold for the CO emission from the
Cloverleaf QSO (\cite{bar94}); thus the emission in the unlensed radio
galaxy 53W002 is more than a factor of four weaker than that in the
previously detected high redshift systems.

In Figure 2 maps of the CO emission integrated over velocity ranges with
$\Delta V$=235 \kms\ are shown. In Figure 3 the total CO line flux
integrated over 200 MHz (590 \kms) centered on the line is shown
superposed on the HST V-band image from Windhorst, Keel, \& Pascarelle
(1997).  The flux in the integrated CO map is $1.51\pm0.2$ Jy \kms\ and
the peak flux in the channel maps is $4.06\pm0.89$ mJy beam$^{-1}$.  No
other CO emission features were detected within our 70 \arcsec ~primary
beam. We can therefore set upper limits at $\leq 1.5$ mJy (2$\sigma$) for
the line flux at the positions of z=2.4 candidates 5, 11, etc of
Pascarelle \etal (1996b).

The deconvolved size of the emission feature seen in the integrated map
(Figure 3) is $5.7\pm1.3$\arcsec~by $1.7\pm0.7$\arcsec\ (FWHM) with major
axis at PA=114\deg.  At the angular size distance of 1.08 Gpc, the major
axis diameter corresponds to 29.9 kpc, or a radius of 15 kpc.  In view of
the limited signal-to-noise ratio in the maps, these size estimates are
highly uncertain; however, it is clear that the emission is resolved since
the centroid is shifted in the channel maps (see Figure 2) with positive
velocity emission (relative to systemic) seen predominantly in the
southeast and negative velocities in the northwest. The fact that this
velocity gradient is along the major axis of the intensity distribution is
suggestive of rotation, possibly in a forming disk.

The peak flux in the narrow band maps (4 mJy) implies a beam averaged
brightness temperature excess of 39 mK at $\lambda$=2.9 mm or 0.13 K in
the rest frame of the source.  This corresponds to an absolute brightness
temperature of 9.5 K in the rest frame after adding in the cosmic
background radiation. No emission, above that expected due to the CO line,
was detected in the 1 GHz (BW) continuum filter at a level $\sigma =
200~\mu$Jy -- the CO line contribution averaged over 1 GHz is expected to
be $\sim 0.5$ mJy.

In Figure 4 the radio spectral energy distribution is shown, including the
lower frequency fluxes of W91 and our measured flux at 15 GHz ($2.8\pm0.2$
mJy) and the measured continuum at 101.9 GHz corrected for the known line
flux. A power law fit to the radio continuum (dashed line) gives a
spectral index $\alpha = 1.2\pm0.1$. The line flux, averaged over 200 MHz
(2.8 mJy) is shown as the solid dot to indicate that it far exceeds the
flux expected from extension of the cm-wavelength radio continuum, and
hence it must be line emission. We also note that there is no evidence at
15 GHz for a significant thermal nuclear component setting in at higher
frequencies since this point lies on the power-law extrapolation of the
lower frequency data. Scaling Arp 220 for the H2 mass and luminosity
distance, the expected thermal dust emission for 53W002 at 102 GHz (345
GHz rest frame) is about 0.1 mJy, and the free-free emission would be
10-100 times less (c.f. Scoville \etal 1997).

\section {Analysis and Discussion} 

The center frequency of the CO(3--2) emission is 101.86 GHz, corresponding
to a mean redshift of z=2.394$\pm$0.001.  This is marginally different
(350 \kms) from the mean redshift z=2.390$\pm$0.001 for the optical/UV
emission lines (W91, Pascarelle \etal 1996a, b).  A similar offset with
higher redshifts for the radio versus optical lines has been noted for
lower redshift, high luminosity galaxies (\cite{ms89}). The offset might
be explained if molecular gas and dust gas is falling into the galaxy, or
if the optical/UV lines originate from gas flowing out of the galaxy with
the symmetric redshifted emission extincted by dust within the system.

{}From the integrated line flux of 1.51$\pm$0.2 Jy \kms, we
obtain a total CO (3-2) luminosity of
L$^\prime_{CO}$=1.86$\times$10$^{10}$ K km s$^{-1}$ pc$^2$, using equation
(3) from Solomon et al. (1992b).  (This is a factor of 3 weaker than that
of FSC10214+4724.) For a Galactic conversion factor $\alpha$=4 M$_\odot$
(K km s$^{-1}$ pc$^2$)$^{-1}$, the total molecular mass is
7.4$\times$10$^{10}$\Msun.  The limits on the mass from other objects
within the primary beam (eg. objects 5, 11, etc) are $\leq
3\times$10$^{10}$\Msun ($2\sigma$). If the clouds are hotter or denser
than Galactic GMC's, as would be expected in a vigorous starburst galaxy,
the conversion factor should vary as ${n^{1/2}}/{T_{CO}}$. To some extent,
the decrease in the conversion factor due to the hotter gas may be offset
by increased density. In Arp 220, an independent dynamical constraint on
the conversion factor suggests a value $\sim 0.45$ times the Galactic
value (\cite{sco97}).

Based on the measured sizes and line-widths, it is also possible to
estimate the dynamical mass associated with the CO emission feature.
Assuming the galaxy is rotating, $M_{dyn} = {RV_{1/2}^2}$/$ {G sin^2 i}$.
For $V_{1/2}$, we adopt 200--250 \kms, i.e. approximately half of the
measured line width, and we take R$=10$--$15$ kpc; we then obtain
$0.9$--$2.2\times$10$^{11}$ M$_\odot$ for the galaxy, with no correction
for inclination. The molecular gas mass, assuming the Galactic CO-to-H$_2$
conversion factor, would therefore constitute between 30--80$\%$ of the
total dynamical mass.

In present epoch galaxies, the gas mass fraction is typically $\leq 10\%$;
the probably much higher values indicated for 53W002 are consistent with
this being a genuinely young galaxy, still in its initial phase of star
formation. In nearby ultraluminous IRAS galaxies, high gas mass fractions
are also indicated (cf. \cite{sco97}), but in these cases the gas is
concentrated almost entirely in the central kpc -- not over 30 kpc as in
53W002.  Based on rest frame UV data for 53W002, W91 and Windhorst \etal
(1997) have derived star formation rate of 50-100 \Msun yr$^{-1}$ in order
to account for the fluxes and colors. The age of the stellar population is
estimated by them to be 0.3-0.5 Gyr in the center and 0.5-1 Gyr at
$\sim10$ kpc radius. At these rates, the estimated mass of molecular gas
would be cycled into stars within $\sim10^9$ yrs; this is also consistent
with a young age for the system.

53W002 also has a weak extension in the optical to the West ($\sim 4$ kpc;
\cite{win97}) which may be seen in the HST B- and V-band images (see 
Figure 3). The east-west orientation of this feature is similar to
that of the radio source (cf. \cite{win91}) and the CO emission  
although the latter is extended on both sides of 53W002. 

Yamada \etal (1995) have reported detection of a possible CO (1-0) feature
from 53W002 at a central redshift z = 2.392 with peak and integrated
fluxes of 5 mJy and 1.92 Jy \kms; however, it seems unlikely that this is
the CO (1-0) counterpart of the CO (3-2) emission reported here in view of
the much lower mass implied by our CO (3-2) flux (and the marginally
different redshift). For optical thick CO emission lines with the same
brightness temperature in both CO transitions, the (3-2) flux should be 9
times greater (if the lines are optically thin, the ratio should be higher
still). For low excitation GMC's in the Galactic disk, the line
temperature ratios are typically (1-0)~:~(2-1)~:~(3-2) = 1~:~0.7~:~0.4
(\cite{san97}). Thus even in the case of optically thick Galactic clouds,
our line flux is approximately a factor of 5 weaker than what would be
expected from the Yamada \etal feature, and in the more likely case of
nearly equal brightness temperatures, it is a factor $\sim$14 weaker.
Lastly, it is possible that Yamada \etal have detected additional flux
from the other 18 z=2.4 objects of Pascarelle \etal (1996b) in the
vicinity of 53W002; however, our maps show no other comparably strong
features within an area larger than their 50\arcsec ~diffraction beam.  CO
emission was not detected by us from the other objects in the
interferometer field of view (70\arcsec) at levels less than 1/3 of that
seen in 53W002. These systems must have less massive ISM's than the galaxy
associated with 53W002. The CO luminosity of 53W002 is approximately three
times that of the nearby ultraluminous IRAS galaxy Arp 220 (cf.
\cite{sco97}) and is similar to the most gas-rich local systems such as
IR14348-14 (\cite{san91}) and the radio galaxy PKS1345+12 (\cite{ms89},
\cite{eva96}).

The large gas-mass found for 53W002 suggests that this is a very gas-rich
and massive young galaxy -- perhaps the progenitor of an elliptical or
early-type spiral galaxy. In the course of its subsequent dynamical
evolution, the gas in 53W002 may concentrate in the center (through
dissipative energy loss as is likely to have happened in the nuclei of the
nearby ultraluminous IRAS galaxies, cf. \cite{sco94}).  The central mass
density could then become high enough to place the system in the area of
the fundamental plane occupied by present-epoch elliptical galaxies
(\cite{ks92}). Already, the light profile in the envelope at radii 6--9
kpc apparently follows an $r^{1/4}$ law (Windhorst \etal 1992, 1994b,
\cite{win97}), for which W91 derived a luminous mass of
$2-4\times10^{11}$\Mo~from nine-band photometry combined with its
spectra.  Lastly, we note that 53W002 is a compact steep spectrum radio
source and at the present epoch, such radio sources are hosted usually by
elliptical galaxies.

In conclusion, we may be witnessing the birth of a luminous elliptical or
early-type spiral galaxy. The copious amount of CO (and $H_2$) gas
detected with OVRO suggest that such formation may have initially been
associated with a massive disk, that is subsequently destroyed by mergers
or tidal effects, eventually to leave a luminous bulge-dominated galaxy.

\acknowledgements

The Owens Valley millimeter array is supported by NSF grants AST 93-14079
and AST 96-13717.  The HST observations were supported by NASA grants
GO-5308.*.93A and GO-5985.*.94A (to RAW and WCK) from STScI, which is
operated by AURA, Inc., under NASA contract NAS5-26555. N. Scoville thanks
Don Hall and Dave Sanders for hospitality at the Institute for Astronomy,
University of Hawaii when part of this work was done. We also thank Aaron
Evans for comments and suggestions on the manuscript.

\clearpage

\begin{deluxetable}{lll}
\tablecaption{Summary of CO Observations and Derived Properties}
\tablehead{
\colhead{}               & 
\colhead{}               & 
\colhead{}      }  
\startdata
RA (B1950) & 17$^h$12$^m$59$s\atop .$86 & \nl
DEC (B1950) & +50$^\circ$18$^\prime$51$^{\prime\prime}_.$3 & \nl
$< z >_{CO}$ & $2.394\pm0.001$ & \nl
Luminosity Distance\tablenotemark{a} & 12.41 Gpc & \nl
Angular-Size Distance\tablenotemark{a} & 1.08 Gpc \hskip 0.3in (1$^{\prime\prime} \rightarrow$ 5.24 kpc) & \nl
$M_{H_2}$\tablenotemark{b} & $(7.4\pm1.5)\times10^{10}$ M$_\odot$ & \nl
$M_{dyn}sin^2i$            & $(9-22)\times10^{10}$ M$_\odot$ & \nl

\tablenotetext{a}{H$_o$ = 75 km s$^{-1}$ Mpc$^{-1}$, q$_o$ = 0.5}
\tablenotetext{b}{using $\alpha$=4 M$_\odot$ (K km s$^{-1}$ pc$^2$)$^{-1}$}
\enddata
\end{deluxetable}

%
%

\def\aas     {A\&AS}

\clearpage

\figcaption[fig1.ps]{Average spectrum of CO (3--2) emission in
53W002 smoothed to 80 MHz (235 \kms) resolution
is shown for a 5\arcsec\ aperture centered on the position of 53W002.
\label{fig1}}

\figcaption[fig2.ps]{Maps of the CO emission in 53W002
averaged over 235 \kms\ velocity ranges are shown as a function of 
coordinate offset from the phase center.  The cross marks the position 
of the 15 GHz continuum peak (see Table 1).  The contours are $-$3, 
$-$2, +2, +3, +4, \& +5 times 0.89 mJy beam$^{-1}$ ($1\sigma$), and 
the natural weighted beam is shown.  The velocities are relative
to z=2.394.  No significant continuum emission
was detected (see \S3).
\label{fig2}}

\figcaption[fig3.ps]{The CO (3--2) emission is shown integrated over all
channels containing significant line emission (v=$-$270 to
+270 km s $^{-1}$).  On the right panel, the $4\times 4''$ area
immediately surrounding the radio continuum peak (cross) is
shown zoomed in, with the CO emission contours superposed
on the HST V-band image (Windhorst \etal 1997).  The CO contours 
correspond to
$-3$, $-2$, +2, +3, +4, and +5, times 0.44 mJy beam$^{-1}$ ($1\sigma$).
\label{fig3}}

\figcaption[fig4.ps]{The radio continuum spectrum of 53W002 is shown with 
points at 101.9 GHz to indicate the measured peak CO (3-2) line flux
(solid dot = 2.8 mJy averaged over 200 MHz), the flux measured in the 1
GHz BW continuum filter and the latter corrected for the CO line flux
averaged over the 1 GHz filter. The last, shown as the solid triangle,
provides an upper limit to the true continuum at 101.9 GHz (i.e.
0.3$\pm0.2$ mJy).  The radio fluxes are from Windhorst \etal (1991) (600
MHz, 1.4 GHz, 8.4 GHz) and this article (15 GHz). The dashed line is a
power law with $\alpha = 1.2$ fit to the cm-wave data; the measured narrow
band line flux at 101.9 GHz is far above the extrapolated radio
continuum.  \label{fig4}}

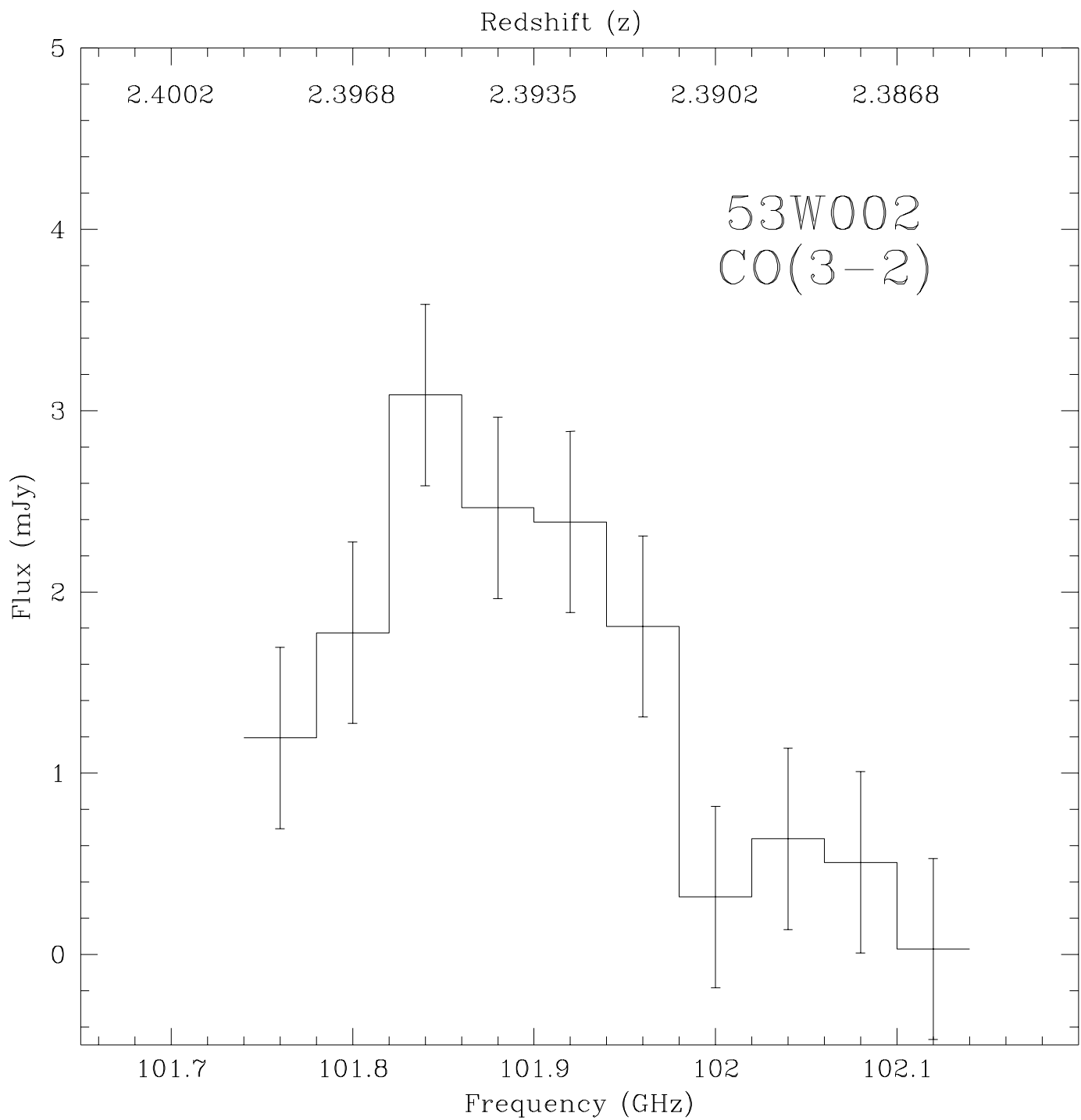
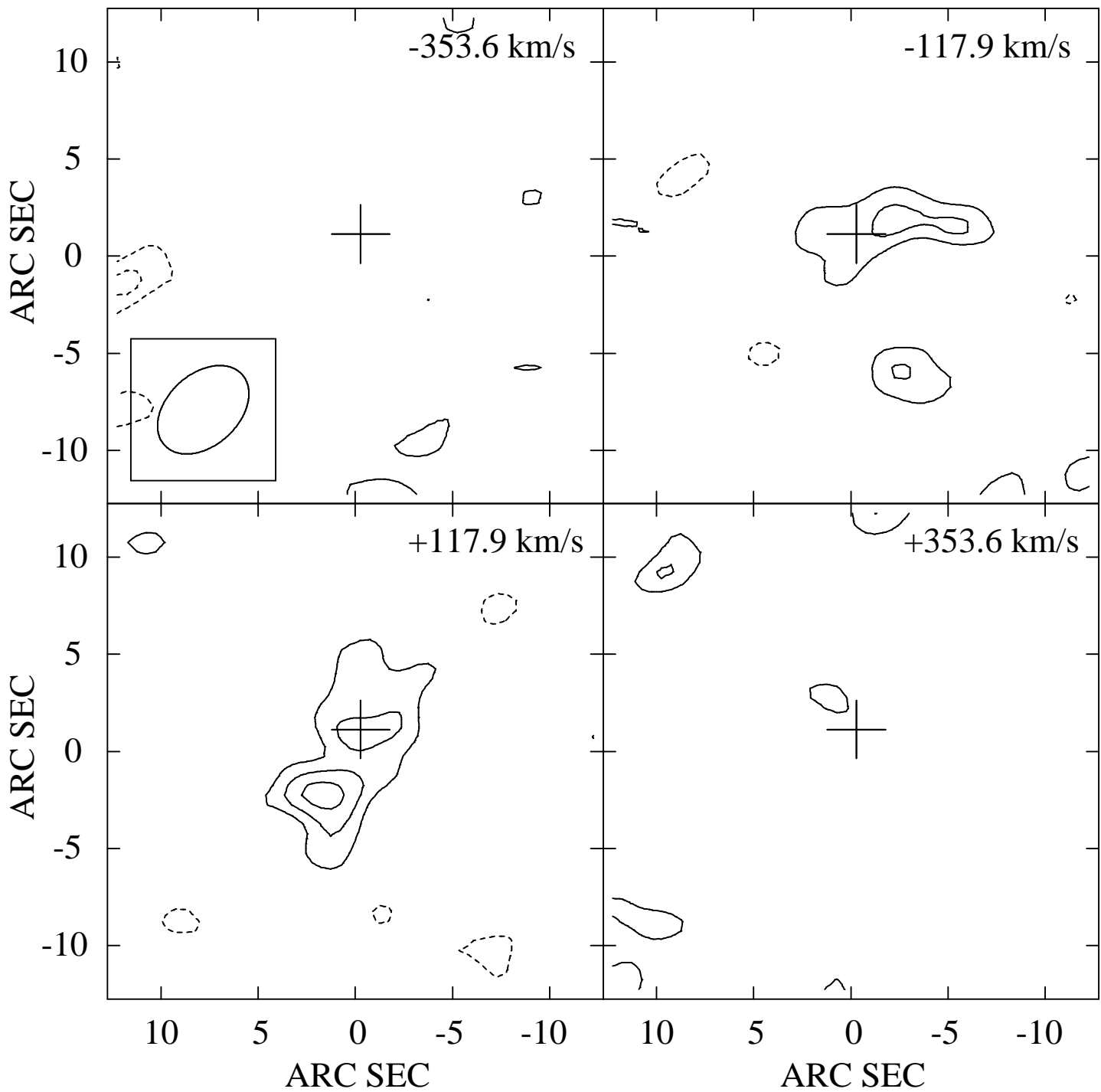
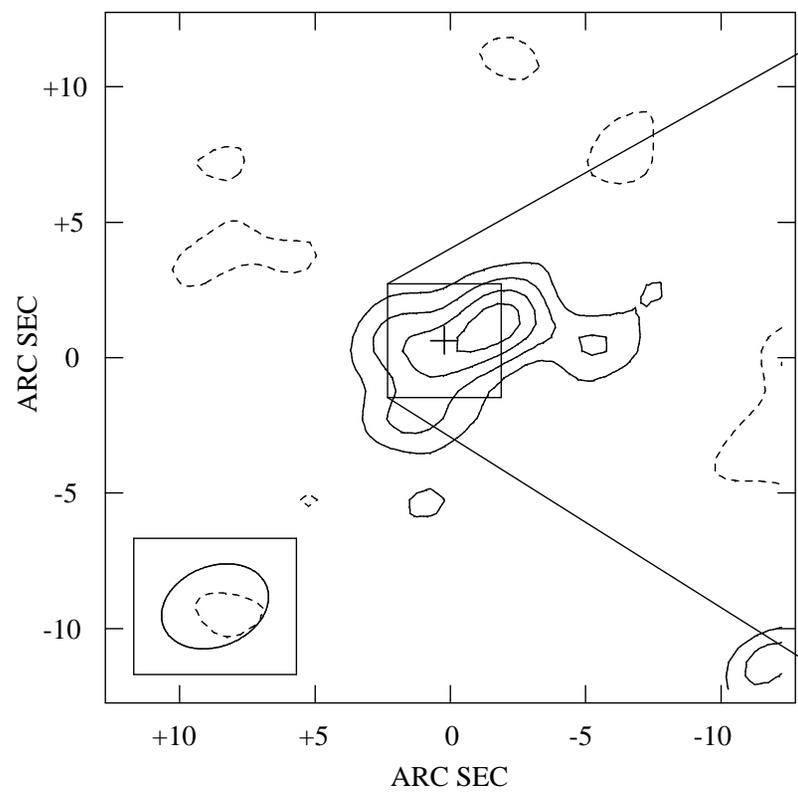
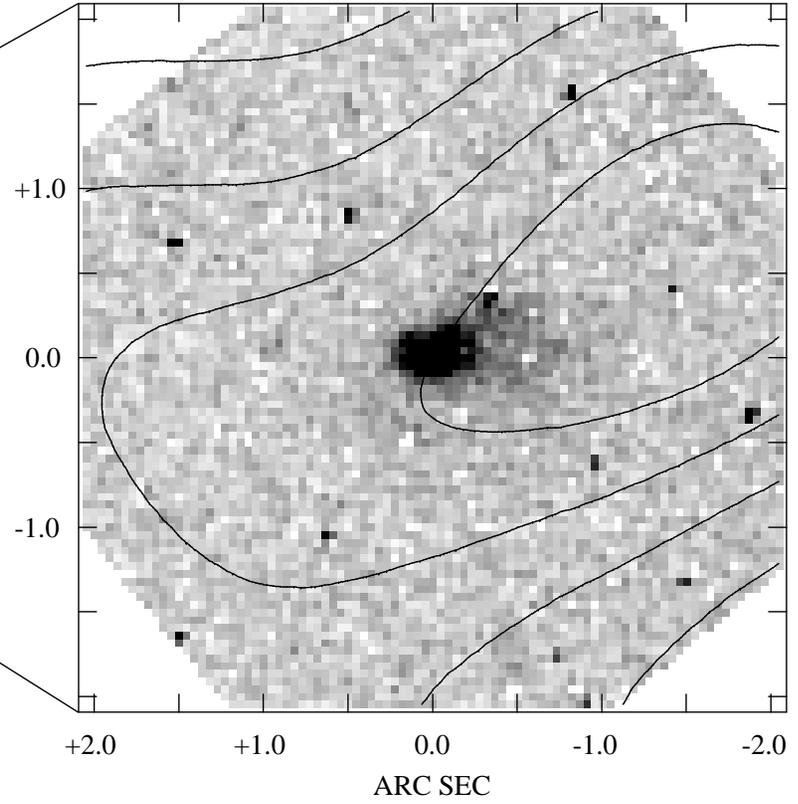
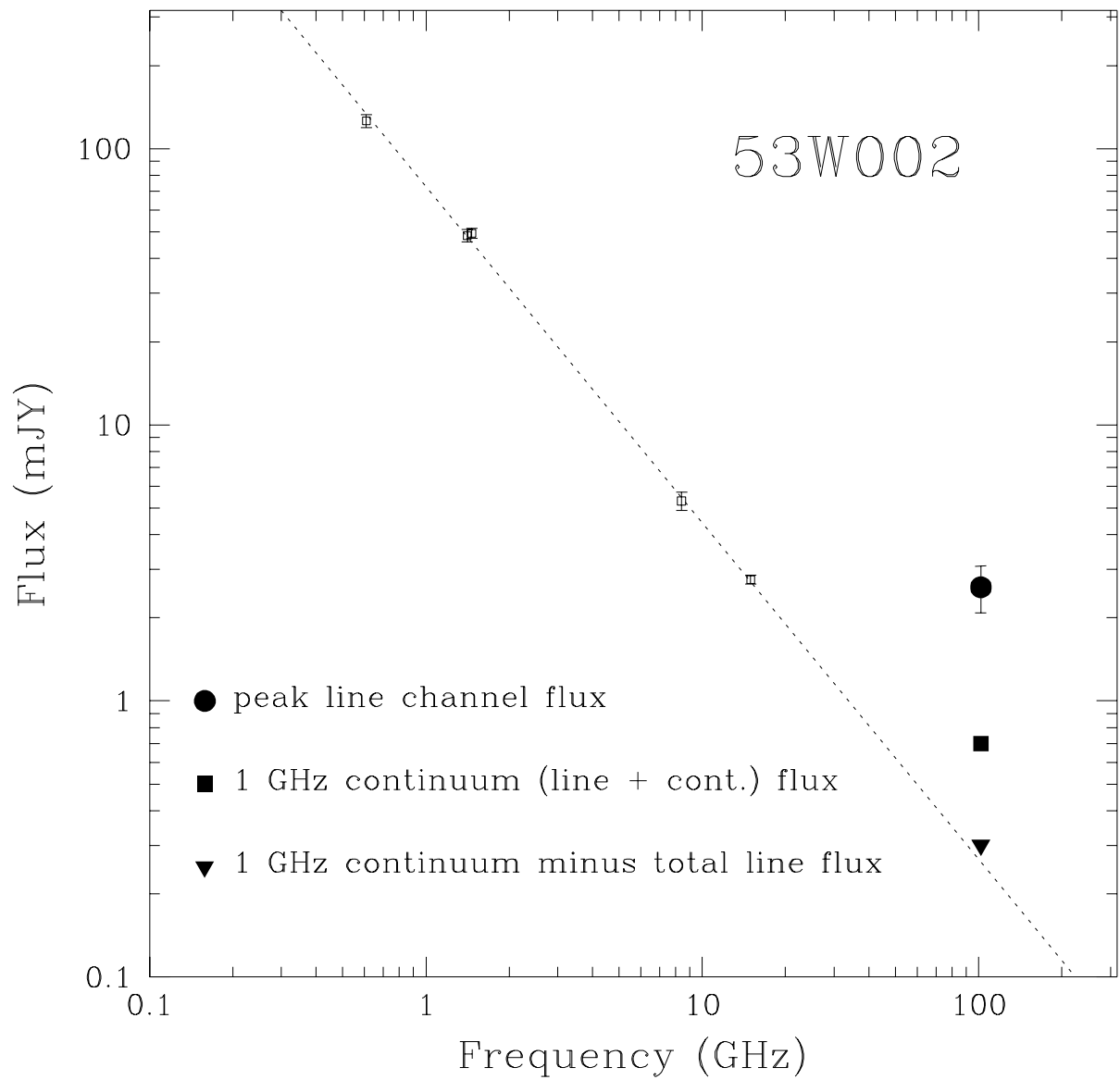
\end{document}